\documentclass[journal=langd5,manuscript=article,layout=twocolumn]{achemso}

\usepackage{style-ftc}
\usepackage[version=3]{mhchem} 
\usepackage[fontsize=10pt]{fontsize}
\usepackage{booktabs}
\usepackage{multirow}
\usepackage{pdfpages}

\let\oldmaketitle\maketitle
\let\maketitle\relax

\author{Fernando Temprano-Coleto}
\affiliation{
    Department of Mechanical and Aerospace Engineering, Princeton University, Princeton, NJ 08544, USA
}
\alsoaffiliation{
    Andlinger Center for Energy and the Environment, Princeton University, Princeton, NJ 08544, USA
}
\author{Jeongmin Kim}
\affiliation{
    Department of Mechanical and Aerospace Engineering, Princeton University, Princeton, NJ 08544, USA
}
\author{Marcel M. Louis}
\affiliation{
    Department of Mechanical and Aerospace Engineering, Princeton University, Princeton, NJ 08544, USA
}
\author{Howard A. Stone}
\email{hastone@princeton.edu}
\affiliation{
    Department of Mechanical and Aerospace Engineering, Princeton University, Princeton, NJ 08544, USA
}
\title[Separation efficiency of diffusiophoresis]{Upper bounds on the \red{colloid} separation efficiency of diffusiophoresis}

\begin{document}
\begin{tocentry}

\includegraphics[]{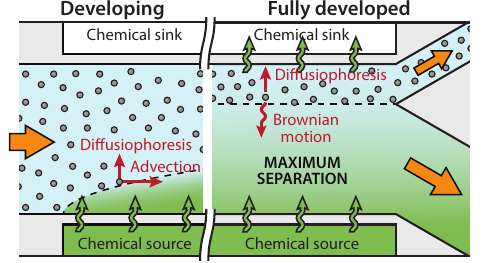}

\end{tocentry}


\twocolumn[
\begin{@twocolumnfalse}
\oldmaketitle
\begin{abstract}
The separation of colloidal particles from fluids is essential to ensure a safe global supply of drinking water, yet in the case of microscopic particles, it remains a highly energy-intensive process when using traditional filtration methods. Water cleaning through diffusiophoresis, spontaneous colloid migration in chemical gradients, effectively circumvents the need for physical filters, representing a promising alternative. This separation process is typically realized in internal flows, where a cross-channel electrolyte gradient drives particle accumulation at walls, with colloid separation slowly increasing in the streamwise direction. However, the maximum separation efficiency, achieved sufficiently downstream as diffusiophoretic migration (driving particle accumulation) is balanced by Brownian motion (inducing diffusive spreading), has not yet been characterized. \red{In this work, we develop an asymptotic theory to predict colloid separation in this limit, deriving expressions for the \emph{water recovery}, defined as the fraction of clean water that can be obtained from the suspension.} We find that the mechanism by which the chemical permeates in the channel and the reaction kinetics governing its dissociation into ions play key roles in the process. Moreover, we identify four distinct regimes in which separation is controlled by different scaling laws involving Damk\"ohler and P\'eclet numbers, which measure the ratios of reaction kinetics to ion diffusion and diffusiophoresis to Brownian motion, respectively. We also confirm the scaling of one of these regimes using microfluidic experiments where separation is driven by $\ce{CO2}$ gradients. Our results shed light on pathways toward new, more efficient separations and are also applicable to quantify colloidal accumulation in the presence of chemical gradients in more general situations.\\\\
%
\vspace*{0.5cm}
\end{abstract}
\end{@twocolumnfalse}
]


\section{Introduction}
\label{sec:mintro}
Colloidal suspensions spontaneously migrate in the presence of electrolyte gradients through a physicochemical mechanism known as diffusiophoresis. Since its discovery by Derjaguin\cite{Derjaguin1947-gq,Derjaguin1961-rw} and its subsequent theoretical formalization \cite{Prieve1982-pp,Anderson1984-zx,Prieve1984-rj,Prieve1987-uj}, many studies have used this phenomenon to manipulate particle motion at small scales. Applications include colloid focusing \cite{Abecassis2008-cw,Shi2016-oj,Liu2025-gw,Sear2026-fb}, custom particle migration \cite{Banerjee2016-ay,Banerjee2019-av,Banerjee2019-kh}, particle delivery into porous structures \cite{Shin2016-pc,Doan2021-ot,Tan2021-fj,Akdeniz2023-zb}, or the characterization of colloidal properties \cite{Shin2017-qm,Rasmussen2020-dg}, among others\cite{Shim2022-rj,Ault2025-wp}.

Diffusiophoresis has also been shown to induce the continuous removal of colloidal particles from water  \cite{Shin2017-lj,Lee2018-go,Shim2021-yt,Shin2020-av,Shimokusu2020-fp,Chakra2025-sr}. In this scenario, separation is typically achieved by field-flow fractionation, where a flowing colloidal suspension is subject to a cross-channel chemical gradient. The resulting phoretic motion drives particles to accumulate near channel walls, creating particle-depleted regions near the channel center that can be diverted to obtain clean water. This chemically driven process requires no electrical input beyond the power required for pumping, and circumvents the use of physical filters that become highly energy intensive for small particles since their membrane pore size must also scale with the particle dimensions\cite{Belfort1994-uq,Tang2011-wc}. Therefore, diffusiophoresis has been highlighted as an energy-efficient alternative to traditional filtration, increasingly needed with the rise of emerging contaminants like microplastics and nanoplastics, now widespread in virtually every source of water\cite{Rochman2018-si,Hale2020-ox,Qian2024-gs}. 

\begin{figure}[!t] 
    \centering 
    \subfloat{\includegraphics[]{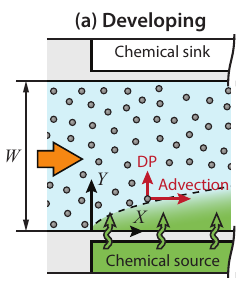}\label{fig:sketch_BL}}
    \subfloat{\includegraphics[]{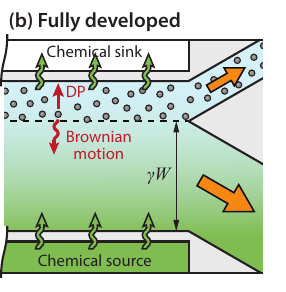}\label{fig:sketch:fully_dev}}
    \caption{Colloid separation through diffusiophoresis in field-flow fractionation. (a) Upstream, the interplay between flow advection and the permeation of the chemical results in a \emph{developing} distributions of chemical species and particles. (b) Sufficiently far downstream, all fields are \emph{fully developed} and invariant in the streamwise direction\red{, yielding the maximum fraction $\gamma$ of clean water (the ``water recovery'') that can be obtained through diffusiophoresis.} }
    \label{fig:sketch}
\end{figure}

Existing theories to predict diffusiophoresis-driven colloid separation have so far focused on how particle distributions evolve downstream after a chemical cross-channel gradient is established\cite{Shim2021-yt,Shimokusu2020-fp} \red{(as in Figure \ref{fig:sketch_BL})}. However, the maximum degree of separation, which is achieved sufficiently downstream as the particle distributions become invariant in the streamwise direction \red{(Figure \ref{fig:sketch:fully_dev})}, has not yet been explored. \red{Such a limit yields an upper bound of the fraction $\gamma$ of clean water, often termed the ``water recovery'', that can be obtained from a particle suspension through diffusiophoresis alone. Predicting this quantity, which we define precisely below, is in turn essential to design channel splittings (Figure \ref{fig:sketch}) optimizing clean water output, and is therefore key to assess the potential scalability and feasibility of this type of separation.}

In this study, we formulate a continuum theory for this regime, where the two dominant effects are the diffusiophoresis-induced advection of particles, driving particles to accumulate at channel walls, and the Brownian motion of the colloids, which diffusively smooths the colloid distribution and hinders separation.

The main conclusion of our analysis is that separation depends strongly on the type of chemical source employed, distinguishing between a gas diffusing into the channel through a nonporous membrane and a liquid permeating through a porous membrane\red{, two distinct configurations that have been realized in microfluidic experiments}. Similarly, dissociation kinetics of the chemical into the ions driving the process also plays a key role, with different predictions depending on the rate of dissociation. Using asymptotic analysis, we provide scaling relations that predict separation in four distinct regimes, which are confirmed to be valid by numerical simulations. Finally, we provide experimental verification of one of these regimes by measuring colloid separation in microfluidic channels under a sustained gradient of carbon dioxide.

\section{Theory}
\label{sec:model}
\subsection{Problem setup}
\label{sec:model_setup}
We consider an aqueous suspension of solid particles flowing in a two-dimensional channel at steady state (Figure \ref{fig:sketch}), with $X$ and $Y$ the streamwise and spanwise coordinates, respectively. The wall at $Y=0$ is a membrane that allows a solute $\sub$ to permeate into the channel from a source, and whose molar concentration we denote by $C_s(X,Y)\defeq\ce{[\sub]}$. This solute dissociates in water through the reaction
\begin{equation} \label{eq:equil}
    \ce{ $\sub$ <=>[k_f][k_r] $\nuc$\,$\cat$ + $\nua$\,$\ani$ },
\end{equation}
with $\cat$ a cation, $\ani$ an anion, $\nuc$ and $\nua$ the number of cation and anion moles produced by the dissociation of one mole of $\mathcal{S}$, and $k_f$ and $k_r$ the forward and reverse rate constants, respectively. \red{Typically, the reverse reaction in \eqref{eq:equil} is neglected for strong electrolytes (e.g., salts like $\ce{NaCl}$, $\ce{KCl}$, $\ce{CaCl2}$, ...), which are considered fully dissociated. However, in weak electrolytes like $\ce{CO2}$, which has been used in experiments involving separation \cite{Shin2017-lj,Shimokusu2020-fp,Shim2021-yt}, both reactions are non-negligible and, as we show below, considering the full dissociation chemistry is key to make quantitative predictions.} We further assume that \eqref{eq:equil} is the dominant equilibrium in the bulk liquid, thus regarding $\ce{[\cat]}$, $\ce{[\ani]}$ and $\ce{[\sub]}$ as much larger than the concentrations of any other species possibly arising from additional chemical reactions. 

This assumption implies that we can consider the solution a \emph{binary electrolyte}, in which case we can define a single ``reduced'' ionic concentration 
\begin{equation} \label{eq:def_ci}
    C_i(X,Y)\defeq\dfrac{\ce{[\cat]}}{\nuc}=\dfrac{\ce{[\ani]}}{\nua}.
\end{equation}
This simplification arises from the electroneutrality of the solution\cite{Newman2012-wk}, which should only break down within distances of the order of the Debye-H\"uckel screening length $\lambda_D$ of the (charged) channel walls. This length typically ranges between a few nanometers and a few micrometers, and is therefore much smaller than the width $W$ of the channel in this work, i.e. $\lambda_D\ll{W}$. We also consider the reactions \eqref{eq:equil} to be elementary and the two species $C_s$ and $C_i$ to be sufficiently dilute to regard the electrolyte as an ideal solution. Accordingly, the reaction rates depend on concentration following the law of mass action, so we define $R \defeq k_f\ce{[\sub]} - k_r\ce{[\cat]^\nuc[\ani]^\nua}$ as an overall chemical reaction rate. Using \eqref{eq:def_ci}, we obtain
\begin{equation}
    R(C_s,C_i) = k_fC_s - k_r\nuc^{\nuc}\nua^{\nua} C_i^\nu,
\end{equation}
where we have defined $\nu\defeq\nuc+\nua$. In the case of symmetric $z$:$z$ electrolytes, such as those arising from the dissociation of salts like $\ce{NaCl}$\cite{Paustian2013-mk} or gases like $\ce{CO_2}$\cite{Shin2017-lj,Shim2020-ny,Shimokusu2020-fp,Shim2021-yt,Shim2021-lf,Shim2022-rj}, we have $\nuc=\nua=1$ and $\nu=2$. However, the value of $\nu$ can be higher for asymmetric electrolytes such as $\ce{CaCl2}$ ($\nu=3$), $\ce{Na3PO4}$ ($\nu=4$), or $\ce{Al2(SO4)3}$ ($\nu=5$), so we assume $\nu\geq{2}$ in general.

The spatial distributions of $C_s$ and $C_i$ are therefore governed by two coupled partial differential equations:
\begin{subequations} \label{eq:pdes_chem_dim}
\begin{align}
    U\pd{C_s}{X}{} &= D_s\left(\pd{C_s}{X}{2}+\pd{C_s}{Y}{2}\right) - R(C_s,C_i),\label{eq:pdes_chem_dim_c} \\[5pt]
    U\pd{C_i}{X}{} &=D_i\left(\pd{C_i}{X}{2}+\pd{C_i}{Y}{2}\right) + R(C_s,C_i),\label{eq:pdes_chem_dim_i} 
\end{align}
\end{subequations}
where $U(Y)$ is the velocity field inside the channel, $D_s$ is the diffusivity of the solute, and $D_i$ is the ambipolar diffusivity\cite{Newman2012-wk} of the ions. The system of equations \eqref{eq:pdes_chem_dim} gives rise to spatial gradients in the ionic concentration, which in turn induces a diffusiophoresis-driven migration velocity
\begin{equation} \label{eq:DP_vel_dim}
\boldsymbol{U}_p = \Gamma_p\dfrac{\boldsymbol{\nabla}{C_i}}{C_i}
\end{equation}
of the particles in the suspension\cite{Prieve1984-rj}. The constant $\Gamma_p$ is the diffusiophoretic mobility, dependent on both the composition of the electrolyte and the properties of the particles, and which can be both positive (leading to ``chemo-attracted'' particles) or negative (resulting in ``chemo-repelled'' particles). \red{We note that Equation \eqref{eq:DP_vel_dim}, while derived in the limit of vanishing double layers compared to the particle size ($\lambda_D/R_p\to{0}$, with $R_p$ the particle radius), admits corrections to $\Gamma_p$ accounting for finite double layer thickness\cite{Shin2016-pc,Gupta2020-oi}. This is especially important in cases where $\lambda_D$ is comparable to the particle size $\lambda_D\lesssim{R_p}$, which can occur for sufficiently small ionic concentrations and particle sizes. In addition, \eqref{eq:DP_vel_dim} is valid for both valence-symmetric and asymmetric binary electrolytes\cite{Gupta2019-oc}. Additional ionic species beyond one cation and one anion preclude the definition of a single ``reduced'' concentration $C_i$ and would lead to more complicated expressions for $\boldsymbol{U}_p$ that lie outside the scope of our analysis.}

Denoting the $X$ and $Y$ components of $\boldsymbol{U}_p$, respectively, as $U_p$ and $V_p$, the concentration $N(X,Y)$ of particles is given by
\begin{equation} \label{eq:pde_part_dim}
    \pd{((U+U_p)N)}{X}{} + \pd{(V_pN)}{Y}{} = D_p\left(\pd{N}{X}{2}+\pd{N}{Y}{2}\right),
\end{equation}
where $D_p$ is the diffusivity of the particles arising from their Brownian motion. After the chemical source is set in contact with the channel at $X=0$ (Figure \ref{fig:sketch_BL}), the solute, ions, and particles are initially \emph{developing}, with $C_s$, $C_i$, and $N$ slowly changing in the streamwise direction $X$. This case, as well as its implications for the separation of colloids, has been studied previously by several authors \cite{Shin2017-lj,Lee2018-go,Shim2021-yt,Shimokusu2020-fp}.

In contrast, here we study the limit in which the system given by \eqref{eq:pdes_chem_dim} and \eqref{eq:pde_part_dim} reaches a \emph{fully developed} state (Figure \ref{fig:sketch:fully_dev}). Sufficiently far downstream, the fields $C_s$, $C_i$ and $N$ become invariant in the streamwise direction, i.e. $\partial_XC_s=\partial_XC_i=0$, leading to
\begin{subequations} \label{eq:odes_chem_dim}
\begin{align}
    D_s\td{C_s}{Y}{2} &= \left[k_f C_s - k_r\nuc^{\nuc}\nua^{\nua}C_i^\nu\right],\label{eq:odes_chem_dim_c} \\[5pt]
    D_i\td{C_i}{Y}{2} &= -\left[k_f C_s - k_r \nuc^{\nuc}\nua^{\nua}C_i^\nu\right].\label{eq:odes_chem_dim_i}
\end{align}
\end{subequations}
%

In this fully developed regime, chemical gradients are sustained through the continuous diffusion of solute into the channel from the chemical source at $Y=0$, but also through its withdrawal into a sink at $Y=W$ (see Figure \ref{fig:sketch:fully_dev}). \red{Accordingly, we set boundary conditions}
\begin{subequations} \label{eq:bcs_chem_dim}
\begin{align} 
    C_s &= \Csmax \quad\text{ at } Y=0, \text{ and} \\
    C_s &= \Csmin \quad\text{ at } Y=W,
\end{align}
\end{subequations}
\red{where $\Csmax$ and $\Csmin$ are reference solute concentrations imposed by the chemical source and sink, respectively.}

\red{We can then estimate the typical ionic concentrations in the channel as the values $\Cimax$ and $\Cimin$ set by $\Csmax$ and $\Csmin$, respectively, at chemical equilibrium. These values can be obtained imposing detailed balance in the reaction term $R(\Csmin,\Cimin)=R(\Csmax,\Cimax)=0$, yielding
\begin{subequations}\label{eq:ref_conc_ion}
\begin{align} 
    \Cimax = \left(\dfrac{k_f\Csmax}{k_r\nuc^{\nuc}\nua^{\nua}}\right)^{1/\nu},\label{eq:ref_conc_ion_0}\\
    \Cimin = \left(\dfrac{k_f\Csmin}{k_r\nuc^{\nuc}\nua^{\nua}}\right)^{1/\nu}.\label{eq:ref_conc_ion_star}
\end{align}
\end{subequations}
}

Boundary conditions for the ionic species $C_i$ depend on the mechanism of solute transport through the channel membrane walls. On one hand, for a \emph{liquid} reservoir the solute itself permeates into the channel via fluid flow through the porous wall. This setup can be achieved, for instance, using hydrogel membranes in microfluidics\cite{Paustian2013-mk,Paustian2015-bb,Shi2016-oj,Nery-Azevedo2017-pq,Shah2022-zp}. In this case, the solution composition at walls must match the \red{equilibrium ionic} concentrations of source and sink, i.e.
\begin{subequations} \label{eq:bcs_ion_liquid_dim}
\begin{align}
    C_i &= \Cimax \quad\text{ at } Y=0, \text{ and} \\
    C_i &= \Cimin \quad\text{ at } Y=W.
\end{align}
\end{subequations}
%

On the other hand, in \emph{gas} sources the chemical permeates through the membrane by way of a solution-diffusion mechanism \cite{Wijmans1995-yn}, then dissolves in the aqueous solution to produce the solute, which in turn dissociates into the ions. This setup has been realized in microfluidics using $\ce{CO_2}$ and gas-permeable polydimethylsiloxane (PDMS) devices\cite{Shin2017-lj,Shimokusu2020-fp,Shim2021-yt}. Since the walls are only gas-permeable, ions in solution cannot diffuse through them, such that
\begin{equation} \label{eq:bc_ion_gas_dim}
    \td{C_i}{Y}{} = 0, \quad\text{ at } Y=0 \text{ and } Y=W.
\end{equation}

Regarding the concentration of particles, in the fully developed regime we expect that $\partial_{X}N=0$, leading to 
\begin{equation} \label{eq:ode_part_dim}
    \td{(V_pN)}{Y}{} = D_p\td{N}{Y}{2},
\end{equation}
which is supplemented by boundary conditions specifying a zero particle flux at channel walls,
\begin{equation} \label{eq:bc_part_dim}
    V_pN - D_p\td{N}{Y}{} = 0, \quad\text{ at } Y=0 \text{ and } Y=W.
\end{equation}
\red{Intuitively, Equation \eqref{eq:bc_part_dim} ensures that any particle flux $V_p{N}$ originating from diffusiophoretic migration is compensated by Brownian diffusion $-D_p\partial_{Y}N$ to yield no net particle flux through either wall. Since the} particle concentration given by \eqref{eq:ode_part_dim} and \eqref{eq:bc_part_dim} is only defined up to a multiplicative constant, \red{we also fix} an average value $N_0$ given by
\begin{equation} \label{eq:integ_part_dim}
    \dfrac{1}{W}\int_{0}^{W}N(Y)\dd{Y} = N_0.
\end{equation}

\textbf{Nondimensionalization.} We seek to understand how the problem parameters affect the spatial distribution of particles $N(Y)$, so we first define the nondimensional variables $y\defeq Y/W$, $c_s\defeq{C_s}/\Csmax$, $c_i\defeq{C_i}/\Cimax$, and $n\defeq N/N_0$
. Following this normalization, the particle concentration follows an ordinary differential equation (ODE) given by
\begin{equation} \label{eq:ode_particles}
    \pm Pe_p\td{}{y}{}\left(v_pn_\pm\right) = \td{n_\pm}{y}{2},
\end{equation}
where we define the particle P\'eclet number as
\begin{equation} \label{eq:def_Pe_p}
    Pe_p \defeq \dfrac{|\Gamma_p|}{D_p} > 0,
\end{equation}
and where the signs distinguish between the case where the mobility is positive $\Gamma_p>0$ and the particles $n_+(y)$ are chemo-attracted, from the case where $\Gamma_p<0$ and the particles $n_-(y)$ are chemo-repelled. We have also defined a nondimensional particle migration velocity $v_p\defeq{V_p}W/\Gamma_p$, related to $c_i(y)$ through 
\begin{equation}\label{eq:DP_vel_nondim}
    v_p = \dfrac{1}{c_i}\td{c_i}{y}{}.
\end{equation}
Equation \eqref{eq:ode_particles}, along with \eqref{eq:DP_vel_nondim} and conditions
\begin{subequations} \label{eq:bcs_particles}
\begin{align} 
    \pm Pe_p v_pn_\pm - \pd{n_\pm}{y}{} &= 0 \quad\text{at }y=0\text{ and }y=1, \label{eq:bcs_particles_flux}\\
    \int_{0}^{1}n_\pm(y)\dd{y} &= 1, \label{eq:bcs_particles_int}
\end{align}
\end{subequations}
which are the nondimensional equivalents of \eqref{eq:bc_part_dim} and \eqref{eq:integ_part_dim}, respectively, can be integrated for an arbitrary ionic concentration $c_i(y)$. The solution is
\begin{equation} \label{eq:sol_particles}
    n_\pm(y) = \dfrac{c_i(y)^{\pm Pe_p}}{\int_{0}^{1}c_i(y)^{\pm Pe_p}\dd{y}}.
\end{equation}
This power-law behavior of the particle distribution in $Pe_p$ has also been recently reported in theories for colloid trapping in chemical gradients\cite{Sear2026-fb}. While Equation \eqref{eq:sol_particles} is an exact result, the particles $n_\pm(y)$ still depend on the concentration $c_i(y)$, which is in turn given by the dimensionless versions of \eqref{eq:odes_chem_dim}, namely
\begin{subequations} \label{eq:odes_chem}
\begin{align}
    \dfrac{1}{Da_s} \td{c_s}{y}{2} &= \left[c_s-c_i^\nu\right],\label{eq:odes_chem_c} \\[5pt]
    \dfrac{1}{Da_i} \td{c_i}{y}{2} &= -\left[c_s-c_i^\nu\right].\label{eq:odes_chem_i}
\end{align}
\end{subequations}
Here, we have defined two Damk\"ohler numbers
\begin{subequations} \label{eq:Da_def}
\begin{align} 
    Da_s &\defeq \dfrac{k_fW^2}{D_s}, \label{eq:def_Da_s}\\
    Da_i &\defeq \dfrac{k_fW^2\Csmax}{D_i\Cimax},  \label{eq:def_Da_i}
\end{align}
\end{subequations}
which indicate the relative rates of the chemical reactions given by \eqref{eq:equil} with respect to diffusion. 

The coupled equations \eqref{eq:odes_chem} can be simplified, first noting that adding \eqref{eq:odes_chem_c} and \eqref{eq:odes_chem_i} leads to
\begin{align} \label{eq:ode_chem_combined}
    \dfrac{1}{Da_s}\td{c_s}{y}{2} + \dfrac{1}{Da_i}\td{c_i}{y}{2} = 0.
\end{align}
We can then differentiate \eqref{eq:odes_chem_i} twice and combine it with \eqref{eq:ode_chem_combined} to obtain
\begin{align} \label{eq:ode_chem_final}
    \td{c_i}{y}{4} = Da_i\td{\left(c_i^\nu\right)}{y}{2} + Da_s\td{c_i}{y}{2},
\end{align}
thereby reducing the coupled system to a single fourth-order nonlinear equation for $c_i$, whose solution requires four boundary conditions to be determined uniquely.  

In the case of \emph{liquid sources}, the boundary conditions are the nondimensional versions of \eqref{eq:bcs_chem_dim} and \eqref{eq:bcs_ion_liquid_dim} for the solute and ions, respectively. Combining those expressions with \eqref{eq:odes_chem_i}, it is possible to cast the four boundary conditions in terms of $c_i$ only, leading to
\begin{subequations} \label{eq:bcs_chem_liquid_nondim}
\begin{alignat}{2} 
    c_i &= 1            \quad&&\text{at } y=0, \label{eq:bcs_chem_liquid_nondim_ci_0}\\
    c_i &= \eps^{1/\nu} \quad&&\text{at } y=1, \label{eq:bcs_chem_liquid_nondim_ci_1}\\
    \td{c_i}{y}{2} &= 0 \quad&&\text{at }y=0\text{ and }y=1, \label{eq:bcs_chem_liquid_nondim_cs}
\end{alignat}
\end{subequations}
where we define $\eps\defeq\Csmin/\Csmax=(\Cimin/\Cimax)^{\nu}$ as the ratio of solute concentration between the sink and the source. Since, in practice, the sink could be engineered to have a very low concentration such that $\eps\ll{1}$, taking $\eps=0$ would seem as a natural simplification. However, in the particular case of liquid sources, a ``perfect'' sink with $\eps=0$ would lead to a zero concentration $c_i(1)=0$ at the wall and, as a consequence, to a divergent particle velocity $|v_p|\to\infty$, according to \eqref{eq:DP_vel_nondim}. This would in turn result in an unbounded distribution $n_-(y)$ of particles in the case in which they are chemo-repelled, highlighting that a small (but nonzero) value of $\eps$ has an effect that is not small in the case of a liquid source. While it is possible to regularize this divergence including other physical effects like a concentration-dependent mobility \cite{Gupta2020-oi}, here we will simply assume a nonzero value $0<\eps<1$ in the case of a liquid source.

In the case of \emph{gas sources}, the boundary conditions are the nondimensional versions of \eqref{eq:bcs_chem_dim} and \eqref{eq:bc_ion_gas_dim}, which can once again be combined with \eqref{eq:odes_chem_i} to arrive at
\begin{subequations} \label{eq:bcs_chem_gas_nondim}
\begin{alignat}{2}
    c_i^\nu - \dfrac{1}{Da_i}\td{c_i}{y}{2} &= 1 \quad &&\text{at }y=0\text{,} \label{eq:bcs_chem_gas_nondim_cs_0}\\
    c_i^\nu - \dfrac{1}{Da_i}\td{c_i}{y}{2} &= 0 \quad &&\text{at }y=1\text{,}\label{eq:bcs_chem_gas_nondim_cs_1}\\
    \td{c_i}{y}{} &= 0 \quad &&\text{at }y=0\text{ and }y=1,\label{eq:bcs_chem_gas_nondim_ci}
\end{alignat}
\end{subequations}
where we have taken $\eps=0$ since, in this case, the limit does not lead to a singular particle velocity because the no-flux condition \eqref{eq:bcs_chem_gas_nondim_ci} prevents $c_i$ from becoming zero at the sink. This means that imperfect sinks with small values $0<\eps\ll{1}$ have only small, higher order effects on the particle distribution in the case of gas sources.

Neither the boundary-value problem for liquid sources, given by \eqref{eq:ode_chem_final} and \eqref{eq:bcs_chem_liquid_nondim}, nor the one for gas sources, given by \eqref{eq:ode_chem_final} and \eqref{eq:bcs_chem_gas_nondim}, are exactly solvable in general. However, we can obtain key insights about the structure of their solutions from an asymptotic analysis\cite{Hinch1991-yu,Bender1999-lf,Leal2007-zv} of the problem in practically relevant physical regimes. To that end, we first  estimate the typical ranges of values of the three dimensionless numbers ($Pe_p$, $Da_s$, and $Da_i$) to identify different distinguished limits in the parameter space.

\textbf{Estimation of parameters.} In order to estimate the P\'eclet number $Pe_p$, the particle diffusivity $D_p$ can be calculated using the Stokes-Einstein relation $D_p=k_BT/(6\pi\mu R_p)$, with $k_B$ the Boltzmann constant, $T$ the absolute temperature, $\mu$ the dynamic viscosity of the medium, and $R_p$ the particle radius. This formula neglects electrostatic effects and is strictly valid only in the limit of a thin double layer compared to the particle size $\lambda_D\ll R_p$, but is nevertheless useful as an order-of-magnitude estimate. For water at room temperature, and for particle sizes in the range $R_p=O(\SI{10}{\nano\meter}-\SI{10}{\micro\meter})$ relevant in practice, we have diffusivities $D_p=O(\SI{e-11}{}-\SI{e-14}{\meter\squared\per\second})$. In turn, at moderate concentrations $\Cimax=O(0.1-\SI{10}{\milli{M}})$ we can expect diffusiophoretic mobilities $\Gamma_p$ to be of the same order of magnitude as the ambipolar diffusivity of the electrolyte\cite{Gupta2020-oi}. Most small ions in aqueous solutions have $D_i=O(\SI{e-10}{}-\SI{e-9}{\meter\squared\per\second})$, so we expect that $Pe_p=|\Gamma_p|/D_p=O(\SI{e2}{}-\SI{e4}{})$. Therefore, we only consider the relevant asymptotic limit of $Pe_p\gg{1}$. This range of values of $Pe_p$ should not come as a surprise, since diffusiophoresis is sought in applications precisely because it overcomes Brownian motion, inducing the directed migration of colloids\cite{Abecassis2008-cw,Shin2016-pc,Tan2021-fj}. 

Estimating the values of the two Damk\"ohler numbers is more challenging due to the wide range of variation of the kinetic constants $k_f$ and $k_r$. Nevertheless, we can distinguish qualitatively between two limiting key regimes, given that the ratio of Damk\"ohler numbers \eqref{eq:Da_def} yields
\begin{equation}
    \dfrac{Da_i}{Da_s} = \dfrac{D_s}{D_i}\dfrac{\Csmax}{\Cimax}.
\end{equation}
Since we can expect $D_s$ and $D_i$ to be of the same order of magnitude, the quantity $Da_i/Da_s$ essentially indicates how displaced is the chemical equilibrium given by \eqref{eq:equil}. If $Da_i\ll Da_s$, then ionic concentrations are dominant, $\Cimax\gg\Csmax$, and the dissociation is \emph{strong}. Conversely, if $Da_i\gg Da_s$, then $\Cimax\ll\Csmax$ and the dissociation is \emph{weak}. \red{For instance, common salts are typically regarded as strong electrolytes, and while their association reaction is typically neglected, some studies report very small values of their equilibrium constants, defined as $K_\text{assoc}=C_{s0}/C_{i0}^2$. For example, for $\ce{NaCl}$ and $\ce{KCl}$, $K_\text{assoc}=O(\SI{e-4}{\meter\tothe{3}\per{mol}})$ [Ref. \citenum{De-Robertis1987-az}], leading to $Da_i/Da_s\approx\sqrt{K_\text{assoc}C_{s0}}=O(\SI{e-2}{})$ at solute concentrations of $C_{s0}=O(\SI{1}{\milli{M}})$. On the other hand, weak electrolytes like $\ce{CO2}$ or $\ce{NH3}$ have dissociation constants $K_\text{dissoc}=C_{i0}^2/C_{s0}$ with values $K_\text{dissoc}=\SI{4.3e-4}{\milli{M}}$ and $K_\text{dissoc}=\SI{5.6e-7}{\milli{M}}$, respectively\cite{Haynes2014-na}, which lead to $Da_i/Da_s=\sqrt{C_{s0}/K_\text{dissoc}}$ with values of $Da_i/Da_s\approx\SI{48}{}$ and $Da_i/Da_s\approx\SI{1.3e3}{}$, respectively, at solute concentrations of $C_{s0}=O(\SI{1}{\milli{M}})$.}

\begin{figure*}[!t]
    \centering 
    \subfloat{\includegraphics[]{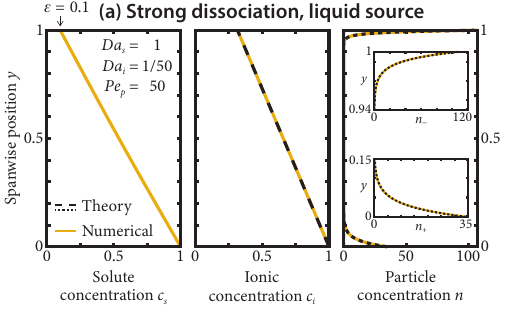}\label{fig:num_sols_strong_liq}}
    \subfloat{\includegraphics[]{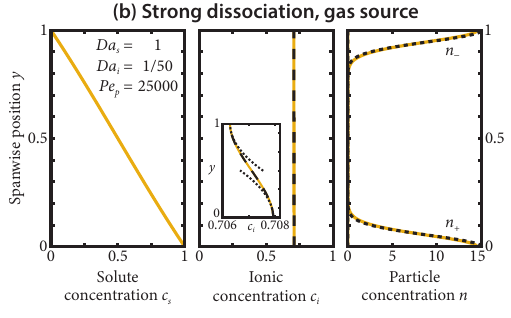}\label{fig:num_sols_strong_gas}} \\
    \subfloat{\includegraphics[]{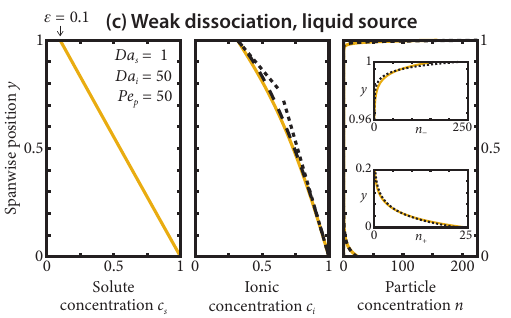}\label{fig:num_sols_weak_liq}} 
    \subfloat{\includegraphics[]{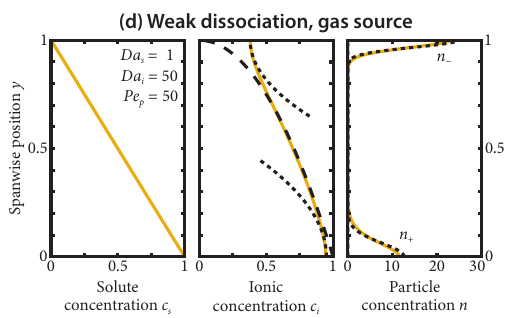}\label{fig:num_sols_weak_gas}}
    \caption{\red{Colloid separation through diffusiophoresis strongly depends on the dissociation chemistry and permeating mechanism of the chemical. Each sub-figure represents one of the four identified characteristic regimes, namely: {\bf(a)} Strong dissociation $[Da_i\ll{Da_s}]$ with a liquid source [Eqs. \eqref{eq:bcs_chem_liquid_nondim}], {\bf(b)} Strong dissociation $[Da_i\ll{Da_s}]$ with a gas source [Eqs. \eqref{eq:bcs_chem_gas_nondim}], {\bf(c)} Weak dissociation $[Da_i\gg{Da_s}]$ with a liquid source [Eqs. \eqref{eq:bcs_chem_liquid_nondim}], and {\bf(d)} Weak dissociation $[Da_i\gg{Da_s}]$ with a gas source [Eqs. \eqref{eq:bcs_chem_gas_nondim}]. All solid lines represent numerical solutions of the full model given by Equations \eqref{eq:sol_particles} and \eqref{eq:ode_chem_final}, subject to boundary conditions given by either \eqref{eq:bcs_chem_liquid_nondim} or \eqref{eq:bcs_chem_gas_nondim}. All dashed lines correspond to asymptotic solutions that are either globally valid or ``outer'' solutions valid only away from walls. All dotted lines correspond to local approximations near walls, obtained either by local expansion of the globally valid solutions or as ``inner'' solutions by asymptotic matching. All insets are magnifications of the solutions for a clearer visualization of their spatial structure.}}
    \label{fig:num_sols}
\end{figure*}

Figure \ref{fig:num_sols} displays the numerical solution of the problem \eqref{eq:ode_chem_final} for $\nu=2$, $Pe_p\gg{1}$, in both the weak and strong dissociation limits, and using either boundary conditions \eqref{eq:bcs_chem_liquid_nondim} for a liquid source (Figure \ref{fig:num_sols_strong_liq}-\subref{fig:num_sols_weak_liq}) or \eqref{eq:bcs_chem_gas_nondim} for a gas source (Figure \ref{fig:num_sols_strong_gas}-\subref{fig:num_sols_weak_gas}). We use $Da_s=1$ in all cases, $Da_i=\red{50}$ for weak dissociation, and $Da_i=\red{1/50}$ for strong dissociation, which are chosen as a representative set of values to capture each regime qualitatively. Additionally, we pick $\epp=0.1$ to regularize the cases with a liquid source and $\epp=0$ for gas sources. Note that the concentration and particles have noticeably different spatial distributions depending on the boundary conditions and the regime of dissociation.

In the next three subsections, we analyze the asymptotic structure of the problem to obtain quantitative predictions for the separation efficiency. We seek to obtain the simplest possible leading-order expression for the particle distributions $n_\pm(y)$ as a function of $Pe_p$, $Da_i$, $Da_s$ (and, in the case of liquid sources, also $\eps$), neglecting all higher order effects.

\subsection{Particle distributions at $Pe_p\gg{1}$}
\label{sec:model_particles}
We first analyze the exact solutions for the distribution of particles given by \eqref{eq:sol_particles}, in the relevant case of $Pe_p\gg{1}$. In this limit, we can expect $n_\pm(y)$ to have a \emph{boundary layer} structure \cite{Hinch1991-yu,Bender1999-lf}, with particles accumulating in thin regions near walls. We denote the width of these regions as $\delta_+$, for the chemo-attracted particles accumulating at the chemical source, and $\delta_-$, for the chemo-repelled particles accumulating at the chemical sink. For high enough P\'eclet number, these regions can get thin enough to approximate the ionic concentration $c_i(y)$ by its leading-order Taylor expansion, since deviations only become noticeable outside the boundary layers $\delta_\pm$. We first introduce local coordinates
\begin{subequations}
\begin{align}
        y_+ &= y \\
        y_- &= 1-y, 
\end{align}
\end{subequations}
to use a more compact notation that ensures $y_\pm=0$ at either wall, and then pose a linear expansion
\begin{equation} \label{eq:taylor_c_i_linear}
    c_i(y_\pm) = c_i(0) + \td{c_i}{y_\pm}{}(0)\,y_\pm + O(y_\pm^2)
\end{equation}
in the case of a liquid source, and a quadratic expansion
\begin{equation} \label{eq:taylor_c_i_quadratic}
    c_i(y_\pm) = c_i(0) + \td{c_i}{y_\pm}{2}(0)\,\dfrac{y_\pm^2}{2} + O(y_\pm^3)
\end{equation}
for gas sources, since in that case we have no ionic flux at walls as given by \eqref{eq:bcs_chem_gas_nondim_ci}. Introducing these expansions into the solution \eqref{eq:sol_particles} for the particles, and after an asymptotic analysis that we detail in the Supporting Information, we can show that, neglecting all exponentially small terms, the leading-order approximations $n^{(0)}$ for the particles are
\begin{subequations} \label{eq:sols_n}
\begin{align}
    n_\pm^{(0)}(y_\pm) &= \dfrac{e^{-y_\pm/\delta_\pm}}{\delta_\pm}\text{ for liquid sources, and} \label{eq:sols_n_exponential}\\
    n_\pm^{(0)}(y_\pm) &= \dfrac{2\,e^{-\big(\tfrac{y_\pm}{\delta_\pm}\big)^2}}{\sqrt{\pi}\,\delta_\pm}\text{ for gas sources.} \label{eq:sols_n_gaussian}
\end{align}
\end{subequations}
The analysis also allows us to identify and define the widths $\delta_\pm$ of the particle boundary layers, which we obtain as
\begin{subequations} \label{eq:bl}
\begin{align}
        \delta_\pm &= \dfrac{c_i(0)}{\left|\td{c_i}{y_\pm}{}(0)\right|}Pe_p^{-1} \text{ for liquid sources,}\label{eq:bl_liquid}\\ 
        \delta_\pm &= \left[\dfrac{2c_i(0)}{\left|\td{c_i}{y_\pm}{2}(0)\right|}\right]^{1/2}Pe_p^{-1/2} \text{ for gas sources,}\label{eq:bl_gas}
\end{align}
\end{subequations}
where we emphasize that the values of $c_i$ and its derivative are evaluated at walls $y_\pm=0$ or, equivalently, at either $y=0$ (for the ``$+$'' case) or $y=1$ (for ``$-$'').

In summary, the particle distributions in the limit of large particle P\'eclet number \red{$Pe_p\gg{1}$} will tend to either exponential \eqref{eq:sols_n_exponential} or Gaussian \eqref{eq:sols_n_gaussian} curves near channel walls, depending on whether the chemical source delivers a liquid or a gas. Furthermore, these distributions and, by extension, the separation of colloids, is dictated by a single boundary layer parameter $\delta_\pm$ at each wall, which itself depends on \red{both the relative importance of diffusiophoresis versus Brownian motion (through $Pe_p$) and potentially on the dissociation chemistry (through the ionic concentration profile $c_i$), as highlighted by Equation \eqref{eq:bl}}. In the next two subsections, we evaluate $c_i(y)$ and their their relevant expansions near walls to obtain $\delta_\pm$ using \eqref{eq:bl}.
\subsection{Strong dissociation ($Da_i\ll{Da}_s$)}
\label{sec:model_strong_diss}
Here, we assume an order-one value of $Da_s=O(1)$, while taking a small $Da_i\ll{1}$ such that $Da_i/Da_s\ll{1}$. Accordingly, we expand $c_i$ in powers of $Da_i$ as
\begin{equation}
    c_i = c_i^{(0)} + Da_ic_i^{(1)} + O(Da_i^2)
\end{equation}
in the governing equation \eqref{eq:ode_chem_final}. At leading-order, we obtain a linear ODE 
\begin{equation} \label{eq:ode_c_i_strong}
    \td{c_i^{(0)}}{y}{4} = Da_s\td{c_i^{(0)}}{y}{2}
\end{equation}
that results in a different ionic profile $c_i^{(0)}(y)$ depending on the specific boundary conditions, as we show next.

\textbf{Liquid sources.} Solving \eqref{eq:ode_c_i_strong} subject to boundary conditions \eqref{eq:bcs_chem_liquid_nondim} results in a linear profile 
\begin{equation} \label{eq:sol_c_i_strong_liq}
    c_i^{(0)}(y) = 1 - (1-\eps^{1/\nu})y,
\end{equation}
which, as displayed in the central panel of Figure \ref{fig:num_sols_strong_liq}, is indistinguishable from the full numerical solution of \eqref{eq:ode_chem_final}. We can then easily obtain the Taylor expansions $c_i(y_+)=1-(1-\epp^{1/\nu})y_+$ and $c_i(y_-)=\epp^{1/\nu}+(1-\epp^{1/\nu})y_-$ at both walls which, when introduced in \eqref{eq:bl}, lead to
%
\begin{subequations} \label{eq:delta_n_strong_liq}
\begin{align}
    \delta_+ &= \dfrac{1}{1-\epp^{1/\nu}}Pe_p^{-1},\\[4pt]
    \delta_- &= \dfrac{\eps^{1/\nu}}{(1-\epp^{1/\nu})}Pe_p^{-1}.
\end{align}
\end{subequations}
The solutions resulting from using \eqref{eq:delta_n_strong_liq} in the profiles given by \eqref{eq:sols_n} are plotted in the right panel of Figure \ref{fig:num_sols_strong_liq}, showing good agreement with the numerical results. The formulas \eqref{eq:delta_n_strong_liq} highlight the \emph{boundary layer} structure, with widths $\delta_+\ll{1}$ and $\delta_-\ll{1}$ that are small since $Pe_p\gg{1}$ and $\eps<1$, which can be confirmed graphically in Figure \ref{fig:num_sols_strong_liq}. This spatial distribution is precisely what is sought to optimize the efficiency of separation and maximize the output of clean water (the filtrate) with respect to the total influx of contaminated water (the feed). In the case of chemo-repelled particles, accumulation is enhanced due to the lower ionic concentration $c_i(1)$ at the chemical sink, which in turn leads to a higher diffusiophoretic velocity. This is apparent from the fact that the expressions for the boundary layer thicknesses \eqref{eq:delta_n_strong_liq} are identical except for an extra factor of $\eps^{1/\nu}<1$, making $\delta_-<\delta+$ and the chemo-repelled particles accumulate more strongly at the top ($y=1$) wall, as shown by Figure \ref{fig:num_sols_strong_liq}. In fact, as $\epp\to{0}$ and the membrane becomes a perfect sink, the boundary layer thickness $\delta_-$ tends to zero and the particle distribution becomes singular, emphasizing the strong effect of $\epp$. Note also that in this particular case the separation does not depend on the reaction chemistry at leading order, given that $\delta_+$ and $\delta_-$ do not depend on $Da_s$ or $Da_i$. 

\textbf{Gas sources.} We proceed by solving \eqref{eq:ode_c_i_strong} with boundary conditions \eqref{eq:bcs_chem_gas_nondim} for gas sources. At leading order in $Da_i$, the solution is $c_i^{(0)}(y)=A$, with $A$ some undetermined constant. This result highlights that this case is characterized by very mild gradients giving a constant ionic concentration and therefore a zero particle velocity at leading order. Physically, this can be understood by noting that the strong dissociation depletes the solute and drives the vast majority of the chemical into the ions which, owing to their no-flux conditions at both walls, tend to a flat profile. The numerical solution plotted in the inset of Figure \ref{fig:num_sols_strong_gas}, however, reveals some very mild gradients, which can only be captured by solving for $c_i^{(1)}(y)$ in the first-order problem. A detailed calculation (see the Supporting Information) leads to
\begin{subequations} \label{eq:sol_c_i_strong_gas}
\begin{align}
    c_i^{(0)}(y) &= 2^{-{1}/{\nu}},\\ 
    c_i^{(1)}(y) &= -\dfrac{\left[y-\dfrac{1}{2}+\dfrac{e^{\sqrt{Da_s}(1-y)}-e^{\sqrt{Da_s}y}}{\sqrt{Da_s}(1+e^{\sqrt{Da_s}})}\right]}{2{Da_s^{1/2}}\tanh\left({\sqrt{Da_s}}/{2}\right)},
\end{align}
\end{subequations}
with the approximate first-order solution $c_i^{(0)}+Da_ic_i^{(1)}$ plotted with a dashed line in Figure \ref{fig:num_sols_strong_gas} and showing excellent agreement with the numerical simulations. It is straightforward to obtain local expansions of $c_i$ at walls, which are now parabolic and displayed with dotted lines in the inset of Figure \ref{fig:num_sols_strong_gas}. Introducing these expansions in \eqref{eq:bl}, we obtain
%
%
\begin{equation} \label{eq:delta_n_strong_gas}
    \delta_+ = \delta_- = 2^{\frac{2\nu-1}{2\nu}}\,Pe_p^{-1/2}Da_i^{-1/2}.
\end{equation}
%
A closer look at \eqref{eq:delta_n_strong_gas} shows that particle distributions are symmetric and, since $Da_i\ll{1}$ in this case, the P\'eclet number must be very large for the boundary layers $\delta_+$ and $\delta_-$ to be small. In fact, a detailed analysis (see the Supporting Information) reveals that $Pe_p\gg Da_i^{-1}$ to have noticeable particle accumulation. Indeed, in Figure \ref{fig:num_sols} all cases display noticeable particle accumulation for a moderately large $Pe_p=50$ except for strong dissociation and gas sources, for which $Pe_p=25000$ to have accumulation comparable to the other three scenarios. This is, once again, a direct consequence of the weak gradients that appear in this case, which strongly limits the feasibility of the separation and suggests that gases with very fast dissociation reactions must be avoided in practice.

\subsection{Weak dissociation ($Da_i\gg{Da_s}$)}
\label{sec:model_weak_diss}
We now keep an order-one value of $Da_s=O(1)$ but take a large $Da_i\gg{1}$ such that $Da_i/Da_s\gg{1}$. Accordingly, we pose an expansion 
\begin{equation}
    c_i = c_i^{(0)} + Da_i^{-1}c_i^{(1)} + O(Da_i^{-2})
\end{equation}
in \eqref{eq:ode_chem_final} which, at leading-order, yields 
\begin{equation} \label{eq:ode_c_i_weak_liq}
    \td{}{y}{2}\left[\left(c_i^{(0)}\right)^\nu\right] = 0.
\end{equation}
This equation can be integrated to obtain the solution $c_i^{(0)}=[K_1+K_2\,y]^{1/\nu}$, with $K_1$ and $K_2$ integration constants. Since, in general, one cannot impose the four required boundary conditions given by either \eqref{eq:bcs_chem_liquid_nondim} or \eqref{eq:bcs_chem_gas_nondim} with only two constants, $c_i$ also has a boundary layer structure. The leading-order equation \eqref{eq:ode_c_i_weak_liq} is therefore only an ``outer'' equation not valid sufficiently close to the walls, where other terms must become non-negligible to ensure all boundary conditions are satisfied. We tackle this boundary layer problem through the method of matched asymptotic expansions \cite{Hinch1991-yu,Bender1999-lf,Leal2007-zv}, with all the details available in the Supporting Information, and outline the results in the next two subsections.

\textbf{Liquid sources.} In this case, two boundary layers of width $l_+=l_-=O(Da_i^{-1/2})$ form at each wall, with two corresponding ``inner'' linear ODEs that can be formulated in these regions. Note the contrast between the \emph{chemical} boundary layers $l_\pm$ for $c_i(y)$ and the particle boundary layers $\delta_\pm$ for $n(y)$, which scale differently. The solution to these inner problems can be matched to the outer solution $c_i^{(0)}=[K_1+K_2\,y]^{1/\nu}$ to obtain the constants, leading to $c_i^{(0)}=[1-(1-\epp)\,y]^{1/\nu}$, which is displayed with dashed lines in Figure \ref{fig:num_sols_weak_liq}. As in previous cases, we only seek to find local expansions of $c_i(y)$ near walls, obtaining
\begin{subequations} \label{eq:local_c_i_weak_liquid}
\begin{align}
    c_i &\approx 1 - \dfrac{1-\epp}{\nu}y_+ \hspace{8pt}\text{at the source,}\\
    c_i &\approx \epp^{1/\nu} + \dfrac{\epp^{\tfrac{1-\nu}{\nu}}(1-\epp)}{\nu}y_- \hspace{8pt}\text{at the sink,}
\end{align}
\end{subequations}
which are plotted as dotted lines in Figure \ref{fig:num_sols_weak_liq}. 
Introducing these expressions in \eqref{eq:bl}, we obtain
\begin{subequations} \label{eq:delta_n_weak_liquid}
\begin{align}
    \delta_+ &= \left(\dfrac{\nu}{1-\epp}\right)Pe_p^{-1},\\[4pt]
    \delta_- &= \left(\dfrac{\nu\epp}{1-\epp}\right)Pe_p^{-1}.
\end{align}
\end{subequations}
These expressions depend on the reaction order (with wider layers for larger $\nu$), but are still independent of kinetics since $Da_i$ and $Da_s$ do not appear in \eqref{eq:delta_n_weak_liquid}. As in the case of a strong dissociation with a liquid source, the boundary layer $\delta_-$ also vanishes as $\epp\to{0}$, although in this case the effect of a small concentration at the sink is stronger given that $\delta_-$ depends on $\epp$ as opposed to $\epp^{1/\nu}$. This is evident comparing the chemo-repelled particle distribution in Figures \ref{fig:num_sols_strong_liq} and \ref{fig:num_sols_weak_liq}, with the latter reaching larger particle accumulation near the sink wall. 

\textbf{Gas sources.} The last case involves a gas source and weak dissociation, with $\ce{CO_2}$-driven diffusiophoresis as a prominent example \cite{Shim2020-ny,Shimokusu2020-fp,Shim2021-lf,Shim2021-yt}. Matched asymptotics reveals that the boundary layer at the source is governed by a linear inner equation with $l_+=O(Da_i^{-1/2})$, while at the sink the inner equation is nonlinear and leads to $l_-=O\Bigl(Da_i^{-\nu/(3\nu-1)}\Bigr)$. The full analysis (see the Supporting Information) yields the local expansions 
\begin{subequations} \label{eq:local_c_i_weak_gas}
\begin{align}
    c_i &\approx 1 - \dfrac{Da_i^{-1/2}}{\nu^{3/2}}-\dfrac{Da_i^{1/2}}{2\nu^{1/2}}y_+^2 \hspace{8pt}\text{at the source,}\label{eq:local_c_i_weak_gas_0}\\[4pt]
    c_i &\approx \alpha Da_i^{\tfrac{-1}{3\nu-1}} + \dfrac{\alpha^\nu}{2}Da_i^{\tfrac{2\nu-1}{3\nu-1}}y_-^2 \hspace{8pt}\text{at the sink},\label{eq:local_c_i_weak_gas_1}
\end{align}
\end{subequations}
plotted with dotted lines in Figure \ref{fig:num_sols_weak_gas}. The constant $\alpha$ cannot be calculated in closed form due to the nonlinear inner problem, but can be easily obtained by numerically solving the boundary-value problem
\begin{subequations} \label{eq:nonlinear_prob_alpha}
\begin{align}
    \td{f}{\xi}{2} &= f^\nu - \xi,\\
    \td{f}{\xi}{} &= 0 \quad\,\,\,\,\,\,\,\,\,\text{at }\xi=0,\\
    f &\to \xi^{1/\nu} \quad\text{as }\xi\to\infty,
\end{align}
\end{subequations}
to then take $\alpha=f(0)$. Note that while the reaction order $\nu$ appears in \eqref{eq:nonlinear_prob_alpha}, the constant $\alpha$ has a very weak dependence on $\nu$, as shown in Table \ref{tab:alpha}.
\begin{table}[b]
\centering
\begin{tabular}{|c||c|c|c|c|}
\hline
$\nu$ & 2 & 3 & 4 & 5 \\ \hline\hline
$-\frac{\nu}{3\nu-1}$ & $-2/5$ & $-3/8$ & $-4/11$ & $-5/14$\\ \hline
$\alpha$ & 0.830965 & 0.839455 & 0.855196 & 0.869162 \\ \hline
\end{tabular}
\caption{Values of the exponent $-\nu/(3\nu-1)$ and the constant $\alpha$ (obtained numerically) appearing in \eqref{eq:local_c_i_weak_gas_1} and \eqref{eq:delta_n_weak_gas_m}, as a function of the reaction order $\nu$.}
\label{tab:alpha}
\end{table}

Introducing the parabolic local expansions \eqref{eq:local_c_i_weak_gas} into \eqref{eq:bl} leads to boundary layer widths given by
\begin{subequations} \label{eq:delta_n_weak_gas}
\begin{align}
    \delta_+ &= \left(2^{1/2}\nu^{1/4}\right)Da_i^{-1/4}Pe_p^{-1/2},\label{eq:delta_n_weak_gas_p}\\[4pt]
    \delta_- &= \left(2^{1/2}\alpha^{\frac{1-\nu}{2}}\right)Da_i^{-\tfrac{\nu}{3\nu-1}}Pe_p^{-1/2}.\label{eq:delta_n_weak_gas_m}
\end{align}
\end{subequations}
In this case, the boundary layers clearly depend on the dissociation kinetics through the Damk\"ohler number, with different power laws in each wall that highlight the problem asymmetry. 

In addition, in the case of chemo-attracted particles we find the additional requirement $Pe_p\gg{Da_i}^{1/2}$ for the model to be valid, since otherwise the particle boundary layer is not contained within the chemical boundary layer (i.e. $\delta_+\ll{l_+}$ is not satisfied), and the underlying approximations of the model break down. In the case of chemo-repelled particles, however, we find that the scalings are such that this situation can never occur, so any value of $Pe_p\gg{1}$ is valid. More details about these conditions can be found in the Supporting Information.  

\subsection{Water recovery}
\label{sec:efficiency}
Equipped with a quantitative theory to predict the accumulation of colloidal particles, we can now compute metrics relevant for colloid separation. We focus on the \emph{water recovery}, typically defined in the literature \cite{Alkhadra2022-go} as the proportion of freshwater that can be obtained from the contaminated feed which, as discussed above, has a (dimensionless) mean particle concentration of 1. Defining the freshwater stream as the region within the channel with particle concentration below a reference value $n(y)<n_\text{ref}$, we can obtain the water recovery $\gamma$ by plugging $n(y_\mathrm{ref})=n_\mathrm{ref}$ into the expressions \eqref{eq:sols_n_exponential} or \eqref{eq:sols_n_gaussian} for the particle distributions, with $y_\mathrm{ref}$ some position where the particle concentration reaches the reference concentration. Then, noting that $\gamma=1-y_\mathrm{ref}$ for chemo-attracted particles, while $\gamma=y_\mathrm{ref}$ if they are chemo-repelled, we can solve for $\gamma$. The result is
\begin{subequations} \label{eq:gamma_N}
\begin{align}
    \gamma &= 1 - \delta_\pm\log\left[\dfrac{1}{\delta_\pm n_\mathrm{ref}}\right] \text{for liquid sources, and}\\[5pt]
    \gamma &= 1 - \delta_\pm\log^{1/2}\left[\dfrac{2}{\sqrt{\pi}\delta_\pm n_\mathrm{ref}}\right]\text{for gas sources.}
\end{align}
\end{subequations}


%


\section{Microfluidic Experiments}
\label{sec:expts}
\subsection{Materials and methods}\label{sec:expts_materials}
We fabricate microfluidic devices using standard soft lithography, where PDMS (Sylgard 184, Dow Corning) with a 10:1 ratio of elastomer base to curing agent is poured over a mold of SU-8 resin (SU-8 2050, Kayaku Advanced Materials) fabricated via photolithography. After curing, the solid polymer is peeled from the mold and bonded to a glass slide where a thin PDMS film has previously been spin-coated and cured. This setup ensures that all channel walls are made of PDMS, preventing any asymmetries in surface chemistry. 

The chip, shown schematically in Figure \ref{fig:expts_setup}, consists of three parallel channels of nominal height and width of $H=\SI{40}{\micro\meter}$ and $W=\SI{250}{\micro\meter}$, respectively. Flows of pure carbon dioxide and nitrogen gas are driven along the two side channels, while a suspension of deionized water (Direct Q3, Millipore) and fluorescent microparticles is injected into the central channel using a syringe pump (Pump 11 Pico Plus Elite, Harvard Apparatus). The channels are separated by walls of thickness $w=\SI{100}{\micro\meter}$ along a segment of length $L=\SI{55}{\milli\meter}$, over which the natural permeability of the polymer to gases \cite{Merkel2000-zg} allows the $\ce{CO2}$ to permeate from the ``source'' channel, dissolve and diffuse across the water, and then permeate again into the flowing nitrogen at the ``sink'', replicating the setup outlined in Figure \ref{fig:sketch} with a monolithic, easy-to-fabricate PDMS chip\cite{Shin2017-lj,Shim2020-ny,Shim2021-lf,Shimokusu2020-fp,Shim2021-yt}. The sustained flow in all three channels ensures a persistent flux of dissolved $\ce{CO2}$ in the spanwise direction, preventing the saturation of the sink with $\ce{CO2}$ over time. 

We note that even though our experiments are three-dimensional (with a channel height $H$ that introduces vertical confinement) and the theory is formulated in two dimensions, they are still comparable. While vertical variations in the flow speed and subsequent Taylor dispersion induces vertical changes in the developing $c_s$, $c_i$, and $n$ profiles, as shown by \citeauthor{Shim2021-yt}\cite{Shim2021-yt}, in the fully developed case streamwise advection is zero and this effect is nonexistent. Furthermore, since the fluxes of solute, ions and particles through the top and bottom walls of the channels are either zero or can be assumed negligible compared to those through the sink and source side walls, we expect the $c_s$, $c_i$, and $n$ profiles to be truly invariant in $x$, provided they have fully developed.

All experiments are performed setting a gauge pressure of $p_{in}=\SI{1.38e4}{\pascal}$ for both $\ce{CO2}$ and $\ce{N2}$ at the channel inlets, which leads to a concentration of $\Csmax\approx\SI{29.5}{\milli{M}}$ of $\ce{CO2}$ dissolved in water at the ``source'' wall (see the Supporting Information). We then expect the $\ce{CO2}$ dissolve in water through the dissociation reaction
\begin{equation} \label{eq:CO2_reaction}
    \ce{CO2 + H2O <=>[k_f][k_r] H+ + HCO3-} ,
\end{equation}
which sets the parameter $\nu=2$ from the reaction stoichiometry. Using well-characterized values of the reaction rate constants\cite{Jolly1991-jc} $k_f$ and $k_r$, we can estimate the scale for the ionic concentration $\Cimax\approx\SI{0.118}{\milli{M}}$ and the Damk\"ohler numbers, as defined in Equation \eqref{eq:Da_def}, $Da_s\approx{1.28}$ and $Da_i\approx{306}$ in all experiments. These values correspond to the ``weak dissociation'' regime, as detailed in the Theory section above, and therefore do not replicate all the different theoretical regimes analyzed. Instead, the experiments focus on validating the specific case of a chemical source with a weakly dissociating gas. Since gas sources with a strong dissociation are predicted by our theory to yield a very low separation efficiency, we expect that the regime in the experiments is the most practically relevant to induce separation through diffusiophoresis using gases. In addition, under these conditions we can estimate the Debye length $\lambda_D\approx\SI{29}{\nano\meter}$, much smaller than any of the channel dimensions or particle sizes used.

\begin{table*}[ht]
\centering
\begin{tabularx}{\textwidth}{|c||*{6}{>{\centering\arraybackslash}X|}}
\hline
Particle type & \multicolumn{2}{c|}{Amine-modified (a-PS)} & \multicolumn{2}{c|}{Carboxylate-modified (c-PS)} & \multicolumn{2}{c|}{Uncoated (PS)}\\ \hline
Diameter [\SI{}{\micro\meter}] & \SI{1}{} & \SI{0.2}{} & \SI{1}{} & \SI{0.1}{} & \SI{1.04}{} & \SI{0.08}{}\\ \hline\hline
Zeta potential $\zeta_p$ [\SI{}{\milli\volt}] & \multicolumn{2}{c|}{\SI{60}{} \red{[Refs. \citenum{Shin2016-pc,Shimokusu2020-fp,Shim2022-tx}]}} & \multicolumn{2}{c|}{\SI{-75}{} \red{[Ref. \citenum{Shimokusu2020-fp}]}} & \multicolumn{2}{c|}{\SI{-50}{} \red{[Refs. \citenum{Shin2016-pc,Shimokusu2020-fp,Shim2022-tx}]}}\\ \hline
Mobility $\Gamma_p$ [\SI{}{\meter\squared\per\second}] & \SI{7.5e-10}{} & \SI{3.2e-10}{} & \SI{-6.1e-10}{} & \SI{-1.0e-9}{} & \SI{-4.6e-10}{} & \SI{-2.9e-10}{} \\ \hline
Diffusivity $D_p$ [\SI{}{\meter\squared\per\second}] & \SI{4.4e-13}{} & \SI{2.2e-12}{} & \SI{4.4e-13}{}& \SI{4.4e-12}{}& \SI{4.2e-13}{}& \SI{5.4e-12}{}\\ \hline
P\'eclet number $Pe_p$ & \SI{1.7e3}{} & \SI{1.5e2}{} & \SI{1.4e3}{} & \SI{2.4e2}{} & \SI{1.1e3}{} & \SI{5.3e1}{}\\ \hline
\end{tabularx}
\caption{Comparison of different particle properties used in the experiments.}
\label{tab:particles}
\end{table*}

We probe suspensions of three different types of particles: amine-modified polystyrene (a-PS), carboxylate-modified polystyrene (c-PS), and uncoated polystyrene (PS). Two different particles sizes are used for each of the three types, with their properties summarized in Table \ref{tab:particles}, allowing us to span a broad range of particle P\'eclet numbers in both chemo-attracted and chemo-repelled colloids. Since the particles size include small diameters below hundreds of nanometers, and since the ionic concentration is relatively low, we compute the mobilities $\Gamma_p$ using the full expressions derived by \citeauthor{Prieve1984-rj}\cite{Prieve1984-rj}, which depend on both the particle size and the electrolyte concentration. The implications of choosing this model for the mobilities, which assumes a binary electrolyte and a constant potential on the particle, are discussed by \citeauthor{Gupta2020-oi}\cite{Gupta2020-oi}. As opposed to the a-PS and PS particles, the c-PS particles increase their $\Gamma_p$ in absolute value as the particle size decreases. The trend, opposite for all other particle types, can occur for highly negatively charged particles due to a subtle interplay between the electrophoretic and chemiphoretic contributions to the mobility\cite{Shimokusu2020-fp}. Particle diffusivities $D_p$ are calculated from the Stokes-Einstein relation\cite{Guazzelli2012-de}. In the experiments with amine-modified particles, we pump a 1 vol\% aqueous solution of 3-aminopropyltriethoxysilane (APTES) for 15 min prior to injecting the article suspension to prevent particle adhesion with the PDMS walls\cite{Shin2020-av,Shimokusu2020-fp}. We use suspensions with a particle volume fraction of $\phi\approx\SI{1.5e-4}{}$ in all experiments.
\subsection{Experimental protocol}\label{sec:expts_protocol}
\begin{figure*}[!ht]
    \centering 
    \subfloat{\includegraphics[]{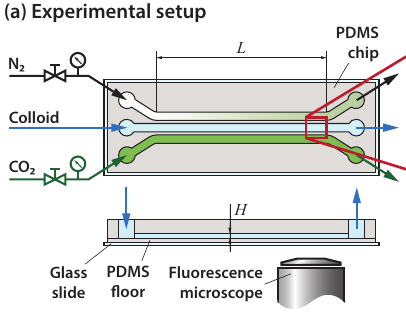}\label{fig:expts_setup}}
    \subfloat{\includegraphics[]{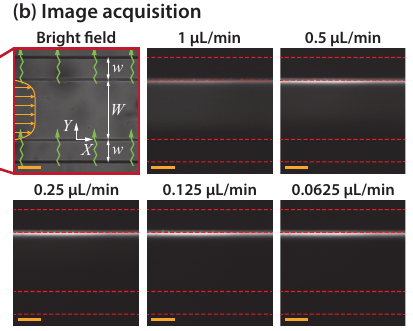}\label{fig:expts_images}}
    \subfloat{\includegraphics[]{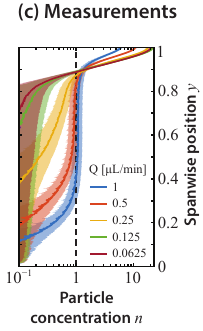}\label{fig:expts_results}} 
    \caption{Microfluidic experiments reproduce the regime of fully developed particle concentrations using $\ce{CO2}$-driven diffusiophoresis. (a) Sketches of the top view and cross-section of the microfluidic devices. (b) Bright-field image of the acquisition window, followed by the fluorescent signal acquired at steady state for five different flow rates. Note that the white particles (in this example, c-PS particles of diameter $\SI{0.1}{\micro\meter}$) progressively accumulate at the top wall (i.e. the sink) with a decreasing flow rate. \red{Scale} bars on the bottom left corner correspond to $\SI{100}{\micro\meter}$. (c) Corresponding fluorescent signal, normalized to yield a mean value of 1 (dashed line). Error bars indicate the standard deviation obtained from averaging the signal in time and in the streamwise direction. The data shows a thin particle-free boundary layer (i.e. an ``exclusion zone'') near the source of $\ce{CO2}$ that grows with a decreasing flow rate. The two curves with the lowest flow rates display a statistically negligible discrepancy, demonstrating that they correspond to the ``fully developed'' regime of interest.}
    \label{fig:expts}
\end{figure*}
We use an inverted fluorescence microscope (Leica  DMI4000B) to image the central channel during a time interval of approximately $\SI{20}{\second}$, at either 10X or 20X magnification (window sizes of 1424$\times$\SI{1064}{\micro\meter} and 712$\times$\SI{532}{\micro\meter}, respectively). A typical acquisition image is shown in Figure \ref{fig:expts_images}, for the case of $\SI{0.1}{\micro\meter}$ diameter c-PS particles. We use the fluorescence signal of the particles (displayed in white) as a proxy for their concentration, ensuring that the signal does not change appreciably during the acquisition time and is therefore in a steady state. For each type and size of particles, we also run a calibration test without $\ce{CO2}$ or $\ce{N2}$ to corroborate that the measured fluorescent signal changes linearly with concentration for the range of particle volume fractions that considered, as detailed in the Supporting Information. 

As shown in Figure \ref{fig:expts_images}, particles accumulate at the (top) sink wall, as expected for this particular case of chemo-repelled particles with a negative mobility $\Gamma_p$. Since our goal is to capture the fully developed regime where particle profiles are invariant in the streamwise $X$ direction, we fix the acquisition window at the end of the channel ($X\approx\SI{55}{\milli\meter}$), and progressively decrease the flow rate to reach the fully developed limit. Note that the flow rates used, shown in Figure \ref{fig:expts_images}, are small (with $Q=0.0625-\SI{1}{\micro\liter\per\minute}$) since experiments are constrained by the typical length of a glass slide (around $\SI{75}{\milli\meter}$). In practically relevant scenarios, channel dimensions could be tuned to maximize flow rate while ensuring that the fully developed regime can be achieved within a feasible channel length, following the analysis in \citeauthor{Shim2021-yt} of the developing particle profiles.

Given that the width of the acquisition windows is much smaller than the channel length $L\approx\SI{55}{\milli\meter}$ over which the particle concentration evolves in the streamwise direction, we average the fluorescence signal in $X$ across the acquisition window, as well as in time. The result, normalized to ensure a unit mean concentration, is plotted in Figure \ref{fig:expts_results}. For the highest three flow rates, the particle profiles are clearly still developing in the streamwise direction: they display a distinct particle free-exclusion zone near the (bottom) source wall, followed by a region of constant particle concentration in the intermediate section of the channel. This constant value is roughly the average $n\approx 1$ incoming from the inlet, reinforcing our interpretation of a developing particle profile with a boundary-layer structure \cite{Shim2021-yt}. In addition, these developing particle profiles display a peak at the (top) sink wall, which we attribute to the dissolved $\ce{CO2}$ naturally present in the water diffusing into the absorbing sink, creating localized gradients that also induce diffusiophoresis. The particle profiles for the lowest two flow rates have no meaningful statistical difference, as evidenced in Figure \ref{fig:expts_results}. This fact confirms that the profile has fully developed and that it can be compared against our theoretical model. The same process of varying the flow rate until a fully developed profile is identified is followed for all experiments, such that only the fully developed profiles are compared to our theory.


\section{Results and discussion} \label{sec:results_discussion}
\begin{figure*}[!ht]
    \centering 
    \subfloat{\includegraphics[]{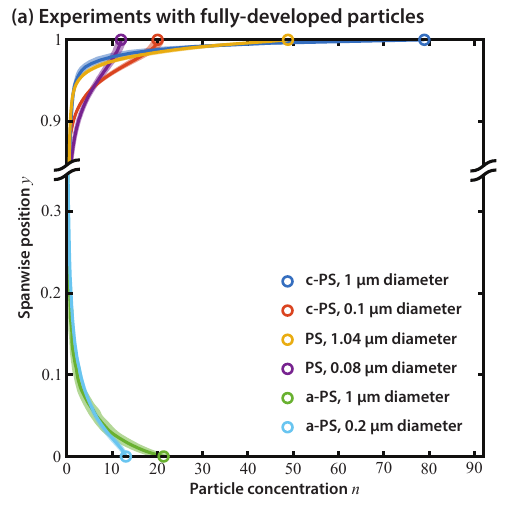}\label{fig:results_profiles}}
    \subfloat{\includegraphics[]{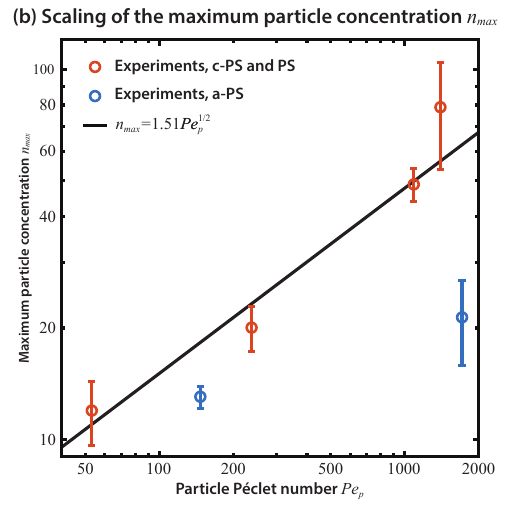}\label{fig:results_scaling}}
    \caption{Experimental measurements of fully developed particle profiles reproduce the scaling predicted by theory. (a) Fluorescent signal, normalized to yield a mean value of 1, for fully developed profiles of particles of different types and sizes, as detailed in Table \ref{tab:particles}. (b) Maximum particle concentration $n_\text{max}$ (at walls $y=\pm{1}$) as a function of the particle P\'eclet number $Pe_p$, as defined in Equation \eqref{eq:def_Pe_p}. Note the trend $n_\text{max}\propto{Pe_p^{1/2}}$, predicted by theory, for chemo-repelled particles (c-PS and PS). For chemo-attracted particles (a-PS), the available experimental data displays significantly lower values of $n_\text{max}$, agreeing qualitatively with theory.}
    \label{fig:results}
\end{figure*}
The fully developed particle profiles obtained in six representative experiments are displayed in Figure \ref{fig:results_profiles}. Each of the six cases corresponds to particles of a different type or size, and is normalized to ensure a unit mean $\int_0^1n\dd{y}=1$. In order to compare the profiles to the model, we focus on the maximum concentration $n_\text{max}=n_\pm(y_\pm=0)$ at the walls, since a metric like the water recovery in Equation \eqref{eq:gamma_N}, while highly relevant in practice, depends on an arbitrary reference particle concentration $n_\text{ref}$ that is difficult to measure experimentally. From Equation \eqref{eq:sols_n_gaussian} and the expressions \eqref{eq:delta_n_weak_gas} for $\delta_\pm$, we expect from theory that $\red{n_\text{max}}=2/(\sqrt{\pi}\delta_\pm)$ and that
\begin{subequations}
\begin{align}
        n_\text{max} &= \dfrac{2^{1/4}}{{\pi}^{1/2}}Da_i^{1/4}Pe_p^{1/2}\text{ at the source,} \\[5pt]
        n_\text{max} &= \left(\dfrac{2\alpha}{\pi}\right)^{1/2}Da_i^{2/5}Pe_p^{1/2}\text{ at the sink.}
\end{align}
\end{subequations}
Qualitatively, we can expect that particles with a higher $Pe_p$ will display higher $n_\text{max}$, and that chemo-attracted particles will achieve lower concentrations than chemo-repelled particles at the same values of $Pe_p$ since, comparing the prefactors, $Da_i^{2/5}>Da_i^{1/4}$. This qualitative trend is reproduced in Figure \ref{fig:results_profiles}, where all particle types show a monotonic increase in $n_\text{max}$ with $Pe_p$, and where the chemo-repelled c-PS and PS particles show higher accumulation than the chemo-attracted a-PS particles.

The scaling $n_\text{max}\propto{Pe_p^{1/2}}$ is also reproduced quantitatively for the four types of chemo-repelled particles, as shown in Figure \ref{fig:results_scaling}. While the availability of only two types of chemo-attracted particles precludes drawing any quantitative conclusion, the measured value of $n_\text{max}$ is also consistently lower than the expected value for chemo-repelled particles of the same $Pe_p$. It is worth noting that the prefactor of approximately 1.5 obtained from fitting the experimental data for chemo-repelled particles deviates from the expected value of $\sqrt{2\alpha/\pi}Da_i^{2/5}\approx{7.1}$ by a factor of approximately 4.7. Such a deviation could originate from complexities in the charge dynamics at the particles, such as charge regulation mechanisms\cite{Shim2022-tx}, from more complex reaction chemistry that is not captured by the single-reaction mechanism \eqref{eq:CO2_reaction}\red{, or due to the effect of particle-particle interactions that introduce dynamics beyond Brownian diffusion\cite{Zaccone2009-bj}}.

The quantitative theory presented here is naturally subject to a number of limitations, the first being that it assumes an elementary dissociation reaction given by Equation \eqref{eq:equil}. While multi-ion electrolytes where concentrations of more than two species are non-negligible would introduce significant differences in the governing equations, complex reaction mechanisms, as long as they can be approximated by an \emph{effective} single reaction of the form \eqref{eq:equil}, would still be captured. Such effective reactions could arise, for instance, with the disparity of different reaction time scales that can be simplified using the well-known quasi-steady-state approximation (QSSA) in chemical kinetics. In fact, the dissociation of carbon dioxide given by \eqref{eq:CO2_reaction} proceeds in two steps $\ce{CO2 + H2O <=> H2CO3 <=> H+ + HCO3-}$, which can be approximated\cite{Kern1960-mj} by the single reaction step given by Equation \eqref{eq:CO2_reaction}.

\red{Similarly, the model includes Equation \eqref{eq:DP_vel_dim} for the diffusiophoretic velocity, which relies on a few assumptions. The first one is that the electrolyte can be regarded as binary, with one cationic and one anionic species being dominant over all other ions, which are assumed negligible in concentration. The expression implicitly relies on the fact that the electrolyte is fully and instantaneously dissociated in the thin electric double layer around the particles\cite{Prieve1984-rj}, which could be violated if the reaction terms \eqref{eq:equil} become important enough to affect the Boltzmann distribution for the ions in that region. A scaling analysis of the governing equations for the ionic concentrations in the double layer, which we detail in the Supporting Information, reveals that Equation \eqref{eq:DP_vel_dim} holds as long as}
\begin{equation} \label{eq:bound_Da_i}
    Da_i \ll \left(\dfrac{W}{\lambda_D}\right)^2.
\end{equation}
\red{For the systems of interest, channel widths are not smaller than $W=O(\SI{100}{\micro\meter})$ and double layers are typically thinner than $\lambda_D=O(\SI{0.1}{\micro\meter})$, which leads to $Da_i\ll{O}(\SI{e6}{})$. We expect this bound to apply for most reactions of practical interest, although there could be extreme cases in which very large $Da_i$ require a different theoretical description.} 

\begin{table*}[ht] 
\centering
\caption{Summary of scaling relations for the nondimensional width $\delta$ of the particle distributions, for all four identified regimes, and conditions of validity of the theory.}
\begin{tabular}{*5c} 
\toprule
{} &  \multicolumn{2}{c}{Liquid source} & \multicolumn{2}{c}{Gas source}\\
\midrule
{}  & Chemo-attracted & Chemo-repelled & Chemo-attracted & Chemo-repelled \\
{}  & particles  & particles   & particles  & particles   \\
\midrule
\multirow{ 2}{*}{\begin{tabular}[t]{@{}c@{}}Strong\\dissociation\end{tabular}} &  $\delta_+=\dfrac{Pe_p^{-1}}{1-\epp^{1/\nu}}$ & $\delta_-=\dfrac{\epp^{1/\nu}Pe_p^{-1}}{1-\epp^{1/\nu}}$   & $\delta_+=2^\frac{2\nu-1}{2\nu}Da_i^{-1/2}Pe_p^{-1/2}$  & $\delta_-=2^\frac{2\nu-1}{2\nu}Da_i^{-1/2}Pe_p^{-1/2}$\\[10pt]
                    &  \begin{tabular}{@{}c@{}}$Pe_p\gg{1}$, \\ $Da_i\redsign{\ll}{1}$\end{tabular}  &  \begin{tabular}{@{}c@{}}$Pe_p\gg{1}$, \\ $Da_i \redsign{\ll}{1}$\end{tabular}   & $Pe_p\gg{Da_i^{-1}}\gg{1}$  & $Pe_p\gg{Da_i^{-1}}\gg{1}$\\
\midrule
\multirow{ 2}{*}{\begin{tabular}[c]{@{}c@{}}Weak\\dissociation\end{tabular}}   &  $\delta_+=\dfrac{\nu Pe_p^{-1}}{1-\epp}$    & $\delta_-=\dfrac{\nu\epp Pe_p^{-1}}{1-\epp}$   & $\delta_+=2^{1/2}\nu^{1/4}Da_i^{-1/4}Pe_p^{-1/2}$  & $\delta_-=2^{1/2}\alpha^\frac{1-\nu}{2}Da_i^\frac{-\nu}{3\nu-1}Pe_p^{-\frac{1}{2}}$\\[10pt]
      &  \begin{tabular}{@{}c@{}}$Pe_p\gg{1}$, \\ $Da_i \gg{1}$\end{tabular}  &  \begin{tabular}{@{}c@{}}$Pe_p\gg{1}$, \\ $Da_i \gg \epp^{\frac{1-3\nu}{\nu}}\gg{1}$\end{tabular}    & $Pe_p\gg{Da_i^{1/2}}\gg{1}$  & \begin{tabular}{@{}c@{}}$Pe_p\gg{1}$, \\ $Da_i \gg{1}$\end{tabular}\\
\bottomrule
\label{tab:scalings}
\end{tabular}
\end{table*}

Our model also assumes a monodisperse distribution of particles, when true colloids (and especially environmental contaminants) are expected to be polydisperse. In the dilute limit $\phi\ll{1}$, particle interactions can be neglected, and we expect that this model could still be easily applicable to a colloidal dispersion with a broad size distribution.

We also note that this theory is only concerned with the limit in which particle distribution is fully developed, i.e., it becomes invariant in the streamwise direction when sufficiently far downstream. Naturally, the downstream length scale over which the particle distribution reaches this fully developed state will also be key in designing separations driven by diffusiophoresis, since it would set a minimum design length to achieve the optimal water recovery given by \eqref{eq:gamma_N}. This length scale, obtained from balancing streamwise advection with cross-channel diffusiophoresis, can be estimated as $L_\text{dev}\sim W^2U/\Gamma_p$, with $U$ a typical velocity scale given by the streamwise flow rate. Further subtleties in the boundary layer analysis, such as the influence of spatially varying shear in rectangular channels, can be found in reference \citenum{Shim2021-yt}.


\section{Conclusions}
We have developed a continuum theory to predict the maximum separation of a colloidal suspension driven by diffusiophoresis in a cross-channel electrolyte gradient. Three governing dimensionless parameters emerge from the theory: two Damk\"ohler numbers $Da_s$ and $Da_i$, given by Equation \eqref{eq:Da_def} and comparing the rates of dissociation kinetics to the diffusion of the solute and resulting ions, and a particle P\'eclet number $Pe_p$, given by Equation \eqref{eq:def_Pe_p} and comparing diffusiophoretic migration to the Brownian diffusion of the particles. The model predicts four distinct regimes \red{(illustrated in Figure \ref{fig:num_sols})} depending on the mechanism by which the delivered chemical permeates into the channel \red{(which we distinguish as either ``liquid'' or ``gas'' sources)} and on whether its dissociation kinetics are ``strong'' ($Da_s\gg Da_i$) or ``weak'' ($Da_s\ll Da_i$). All four regimes are governed by a single parameter $\delta$ measuring the width of the particle distribution, which displays distinct scalings in each regime as a function of $Pe_p$ and, in certain regimes, also of $Da_i$ and of a parameter $\epp$ that accounts for a small concentration at the chemical sink that may arise from imperfect transport through the channel walls. In addition, due to the asymmetry in the ion concentration and the boundary conditions at each wall, we distinguish between $\delta_+$ at the source (for chemo-attracted particles) and $\delta_-$ at the sink (for chemo-repelled particles), which in general have different scalings. \red{From these expressions for $\delta_\pm$, we calculate the \emph{water recovery} $\gamma$ \eqref{eq:gamma_N} as a proxy of the colloid separation efficiency, although the insights about the structure of solutions that we have provided could be easily adapted to calculate other metrics quantifying separation. These metrics can then be used to design channel splittings that maximize the amount of filtrate (clean water) while minimizing the retentate (the particle-rich water stream in Figure \ref{fig:sketch:fully_dev}). All the expressions from} the theory are summarized in Table \ref{tab:scalings}, as well as the conditions of validity of the model. To conclude, we present microfluidic experiments using $\ce{CO2}$-driven diffusiophoresis that confirm the theoretical scaling in one of these four regimes, using polystyrene particles with three different surface functionalization and with diameters ranging from $\SI{0.08}{\micro\meter}$ to \SI{1}{\micro\meter}.

Our analysis provides novel insights into the feasibility of diffusiophoresis as a method for colloid separation. First, it highlights that using a gas that strongly dissociates in water ($Da_i\ll{Da_s}$) leads to an unfavorable regime with low colloid separation, potentially unfeasible in applications. More importantly, it provides quantitative bounds for the separation efficiency in all regimes, essential to assess its potential scalability and design separations driven by diffusiophoresis.


\begin{acknowledgement}
We thank Suin Shim and J. Pedro de Souza for insightful discussions, and Mariko Storey-Matsutani for the assistance developing the experimental protocol. F.T-C. acknowledges support from a Distinguished Postdoctoral Fellowship from the Andlinger
Center for Energy and the Environment. We acknowledge partial support from the National Science Foundation through grant CBET 2127563.
\end{acknowledgement}


\begin{suppinfo} 

Derivation of the leading-order particle profiles from the theory, detailed asymptotics for the ionic concentration profiles, and calibration of particle fluorescence intensity in the experiments.

\end{suppinfo}


\bibliography{paperpile_DP_separation}

    \newpage
    \appendix
    \includepdf[pages=-]{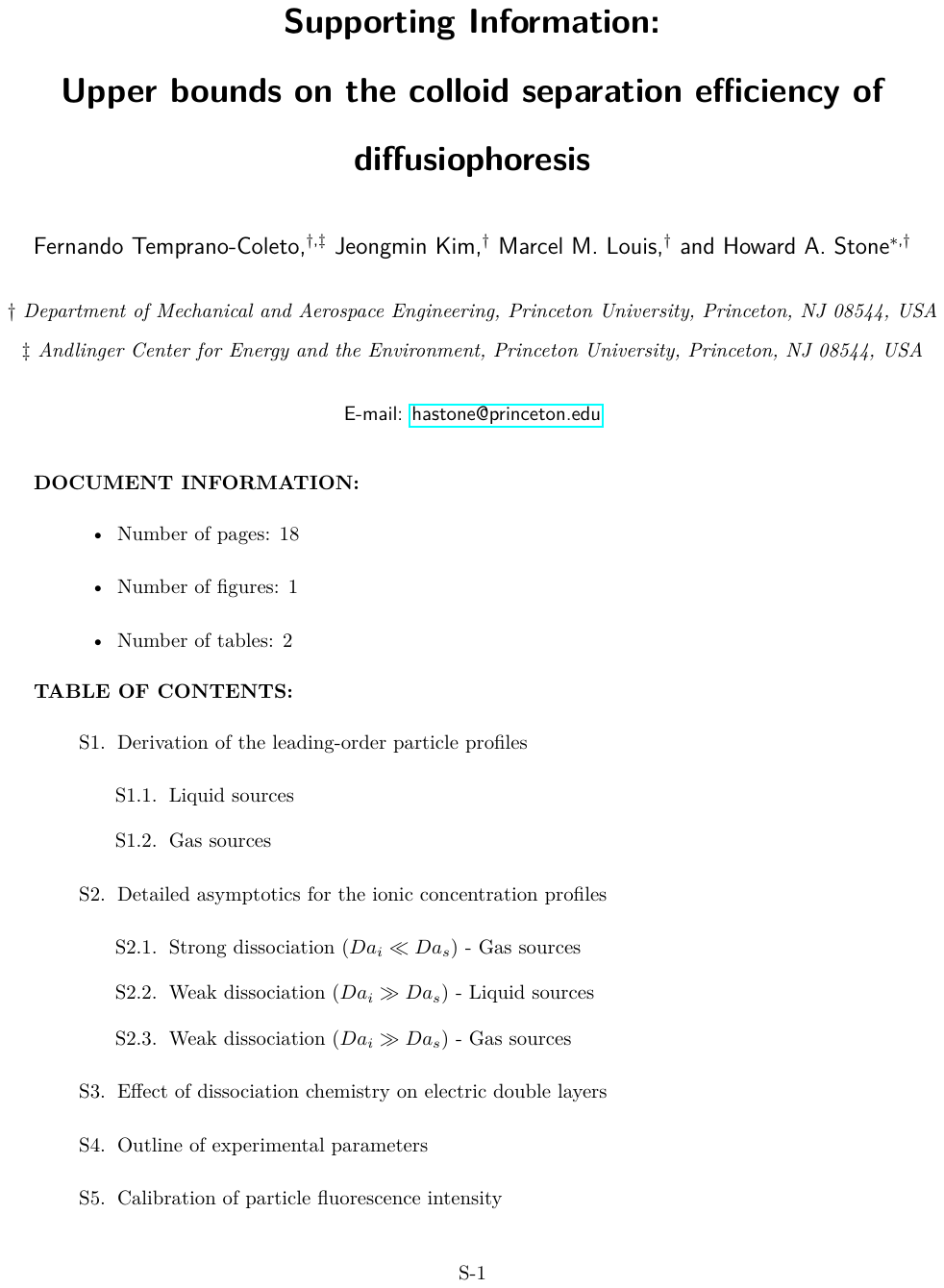}
\end{document}